\documentclass[runningheads]{llncs}
\usepackage{amssymb}
\usepackage{multirow}
\usepackage[T1]{fontenc}
\usepackage{lipsum}
\usepackage[colorlinks,urlcolor=blue]{hyperref}  
\usepackage{color}
\usepackage{graphicx}
\usepackage{float}
\usepackage{subfigure}
\usepackage{soul}
\usepackage[misc]{ifsym}
\usepackage{color}
\begin{document}
%
\title{Contrastive Diffusion Model with Auxiliary Guidance for Coarse-to-Fine PET Reconstruction}
\titlerunning{ }
\authorrunning{ }
%
\author{Zeyu Han\inst{1} \and
Yuhan Wang\inst{1} \and 
Luping Zhou\inst{2} \and 
Peng Wang\inst{1}
Binyu Yan\inst{1} \and 
Jiliu Zhou\inst{1,3} \and 
Yan Wang\inst{1}\textsuperscript{\(\left(\textrm{\Letter}\right)\)} \and 
Dinggang Shen\inst{4,5}\textsuperscript{\(\left(\textrm{\Letter}\right)\)}}

\newcommand\blfootnote[1]{%
  \begingroup
  \renewcommand\thefootnote{}\footnote{\hspace{-1em}#1}%
  \addtocounter{footnote}{-1}%
  \endgroup
}

%
%
\institute{School of Computer Science, Sichuan University, China \\ 
\email{wangyanscu@hotmail.com} \and 
School of Electrical and Information Engineering, University of Sydney, Australia \and
School of Computer Science, Chengdu University of Information Technology, Chengdu, China \and
School of Biomedical Engineering, ShanghaiTech University, China \\ 
\email{dinggang.shen@gmail.com} \and
Department of Research and Development, Shanghai United Imaging Intelligence Co., Ltd.,
Shanghai, China
}

\maketitle              
\blfootnote{Z. Han and Y. Wang -- These authors contributed equally to this work.}

\begin{abstract}

To obtain high-quality positron emission tomography (PET) scans while reducing radiation exposure to the human body, various approaches have been proposed to reconstruct standard-dose PET (SPET) images from low-dose PET (LPET) images. One widely adopted technique is the generative adversarial networks (GANs), yet recently, diffusion probabilistic models (DPMs) have emerged as a compelling alternative due to their improved sample quality and higher log-likelihood scores compared to GANs. Despite this, DPMs suffer from two major drawbacks in real clinical settings, i.e., the computationally expensive sampling process and the insufficient preservation of correspondence between the conditioning LPET image and the reconstructed PET (RPET) image. To address the above limitations, this paper presents a coarse-to-fine PET reconstruction framework that consists of a coarse prediction module (CPM) and an iterative refinement module (IRM). The CPM generates a coarse PET image via a deterministic process, and the IRM samples the residual iteratively. By delegating most of the computational overhead to the CPM, the overall sampling speed of our method can be significantly improved. Furthermore, two additional strategies, i.e., an auxiliary guidance strategy and a contrastive diffusion strategy, are proposed and integrated into the reconstruction process, which can enhance the correspondence between the LPET image and the RPET image, further improving clinical reliability. Extensive experiments on two human brain PET datasets demonstrate that our method outperforms the state-of-the-art PET reconstruction methods. The source code is available at \url{https://github.com/Show-han/PET-Reconstruction}.

\end{abstract}

\section{Introduction}

Positron emission tomography (PET) is a widely-used molecular imaging technique that can help reveal the metabolic and biochemical functioning of body tissues. According to the dose level of injected radioactive tracer, PET images can be roughly classified as standard-(SPET) and low-dose PET (LPET) images. SPET images offer better image quality and more information in diagnosis compared to LPET images containing more noise and artifacts. However, the higher radiation exposure associated with SPET scanning poses potential health risks to the patient. Consequently, it is crucial to reconstruct SPET images from corresponding LPET images to produce clinically acceptable PET images.\\
\indent In recent years, deep learning-based PET reconstruction approaches \cite{gong2018pet,haggstrom2019deeppet,kim2018penalized} have shown better performance than traditional methods. Particularly, generative adversarial networks (GANs) \cite{goodfellow2020generative} have been widely adopted \cite{kaplan2019full,lei2019whole,luo20213d,ouyang2019ultra,wang20183d,wang20183d2} due to their capability to synthesize PET images with higher fidelity than regression-based models \cite{xiang2017deep,xu2017200x}. For example, Kand \textit{et al.} \cite{kang2021translating} applied a CycleGAN model to transform amyloid PET images obtained with diverse radiotracers. Fei \textit{et al.} \cite{fei2022classification} made use of GANs to present a bidirectional contrastive framework for obtaining high-quality SPET images. Despite the promising achievement of GAN, its adversarial training is notoriously unstable \cite{salimans2016improved} and can lead to mode collapse \cite{metz2016unrolled}, which may result in a low discriminability of the generated samples, reducing their confidence in clinical diagnosis. \\
\indent Fortunately, likelihood-based generative models offer a new approach to address the limitations of GANs. These models learn the distribution's probability density function via maximum likelihood and could potentially cover broader data distributions of generated samples while being more stable to train. As an example, Cui \textit{et al.} \cite{cui2022pet} proposed a model based on Nouveau variational autoencoder for PET image denoising. Among likelihood-based generative models, diffusion probabilistic models (DPMs) \cite{ho2020denoising,sohl2015deep} are noteworthy for their capacity to outperform GANs in various tasks \cite{dhariwal2021diffusion}, such as medical imaging \cite{song2021solving} and text-to-image generation \cite{rombach2022high}. DPMs consist of two stages: a forward process that gradually corrupts the given data and a reverse process that iteratively samples the original data from the noise. However, sampling from a diffusion model is computationally expensive and time-consuming \cite{ulhaq2022efficient}, making it inconvenient for real clinical applications. Besides, existing conditional DPMs learn the input-output correspondence implicitly by adding a prior to the training objective, while this learned correspondence is prone to be lost in the reverse process \cite{zhu2022discrete}, resulting in the RPET image missing crucial clinical information from the LPET image. Hence, the clinical reliability of the RPET image may be compromised.\\
\indent Motivated to address the above limitations, in this paper, we propose a coarse-to-fine PET reconstruction framework, including a coarse prediction module (CPM) and an iterative refinement module (IRM). The CPM generates a coarse prediction by invoking a deterministic prediction network only once, while the IRM, which is the reverse process of the DPMs, iteratively samples the residual between this coarse prediction and the corresponding SPET image. By combining the coarse prediction and the predicted residual, we can obtain RPET images much closer to the SPET images. To accelerate the sampling speed of IRM, we manage to delegate most of the computational overhead to the CPM \cite{chung2022come,whang2022deblurring}, hoping to narrow the gap between the coarse prediction and the SPET initially. Additionally, to enhance the correspondence between the LPET image and the generated RPET image, we propose an auxiliary guidance strategy at the input level based on the finding that auxiliary guidance can help to facilitate the reverse process of DPMs, and reinforce the consistency between the LPET image and RPET image by providing more LPET-relevant information to the model. Furthermore, at the output level, we suggest a contrastive diffusion strategy inspired by \cite{zhu2022discrete} to explicitly distinguish between positive and negative PET slices. To conclude, the contributions of our method can be described as follows:


\begin{itemize}
    \item We introduce a novel PET reconstruction framework based on DPMs, which, to the best of our knowledge, is the first work that applies DPMs to PET reconstruction.
    \item To mitigate the computational overhead of DPMs, we employ a coarse-to-fine design that enhances the suitability of our framework for real-world clinical applications.
    \item We propose two novel strategies, i.e., an auxiliary guidance strategy and a contrastive diffusion strategy, to improve the correspondence between the LPET and RPET images and ensure that RPET images contain reliable clinical information.
\end{itemize}


\section{Background: Diffusion Probabilistic Models}
\label{background}
\textbf{Diffusion probabilistic models (DPMs):} DPMs \cite{ho2020denoising,sohl2015deep} define a \textit{forward process}, which corrupts a given image data $x_0 \sim q(x_0)$ step by step via a fixed Markov chain $q(x_t|x_{t-1})$ that gradually adds Gaussian noise to the data:
\begin{equation}
q(x_t|x_{t-1}) = \mathcal{N}(x_t; \sqrt{\alpha_t}x_{t-1}, (1-\alpha_t) I), t=1,2,\cdots,T,
\end{equation}
where $\alpha_{1:T}$ is the constant variance schedule that controls the amount of noise added at each time step, and $q(x_T) \sim \mathcal{N}(x_T;0,I)$ is the stationary distribution. Owing to the Markov property, a data $x_t$ at an arbitrary time step $t$ can be sampled in closed form:
\begin{equation} \label{xt}    
    q(x_t|x_0) = \mathcal{N}(x_t; \sqrt{\gamma_t}x_0, (1-\gamma_t) I); x_t=\sqrt{\gamma_t}x_0 + \sqrt{1-\gamma_t}\epsilon, \epsilon \sim \mathcal{N}(0,I),
\end{equation}
where $\gamma_t = \prod_{i=1}^t \alpha_i$. Furthermore, we can derive the posterior distribution of $x_{t-1}$ given $(x_0, x_t)$ as $q(x_{t-1}|x_0, x_t) = \mathcal{N}(x_{t-1}; \hat{\mu}(x_0, x_t) , \sigma_t^2 I)$, where $\hat{\mu}(x_0, x_t)$ and $\sigma_t^2$ are subject to $x_0$, $x_t$ and $\alpha_{1:T}$. Based on this, we can leverage the \textit{reverse process} from $x_T$ to $x_0$ to gradually denoise the latent variables by sampling from the posterior distribution $q(x_{t-1}|x_0, x_t)$. However, since $x_0$ is unknown during inference, we use a transition distribution $p_\theta (x_{t-1}|x_t):= q(x_{t-1}|{\mathcal{H}}_{\theta}(x_t, t), x_t)$ to approximate $q(x_{t-1}|x_0, x_t)$,
where ${\mathcal{H}}_{\theta}(x_t, t)$ manages to reconstruct $x_0$ from $x_t$ and $t$, and it is trained by optimizing a variational lower bound of $log p_{\theta} (x)$.\\
\textbf{Conditional DPMs:} Given an image $x_0$ with its corresponding condition $c$, conditional DPMs try to estimate $p(x_0|c)$. To achieve that, condition $c$ is concatenated with $x_t$ \cite{saharia2022image} as the input of ${\mathcal{H}}_{\theta}$, denoted as ${\mathcal{H}}_{\theta}(c,x_t,t)$.\\
\textbf{Simplified training objective:} Instead of training ${\mathcal{H}}_{\theta}$ to reconstruct the $x_0$ directly, we use an alternative parametrization ${\mathcal{D}}_{\theta}$ named \textit{denoising network} \cite{ho2020denoising} trying to predict the noise vector $\epsilon \sim \mathcal{N}(0, I)$ added to $x_0$ in Eq.\ref{xt}, and derive the following training objective:
\begin{equation} \label{obj}
    {\mathcal{L}}_{DPM} = \mathbb{E}_{(c, x_0) \sim p_{train}} \mathbb{E}_{\epsilon \sim \mathcal{N} (0, I)} \mathbb{E}_{\gamma \sim p_{\gamma}} \Vert {\mathcal{D}}_{\theta}(c, \sqrt{\gamma} x_0 + \sqrt{1-\gamma} \epsilon,\gamma) - \epsilon \Vert_1,
\end{equation}
where the distribution $p_{\gamma}$ is the one used in WaveGrad \cite{chen2020wavegrad}. Note that we also leverage techniques from WaveGrad to let the denoising network ${\mathcal{D}}_{\theta}$ conditioned directly on the noise schedule $\gamma$ rather than time step $t$, and this gives us more flexibility to control the inference steps.
\begin{figure}[h]
    \centering
\includegraphics[width=0.9\textwidth]{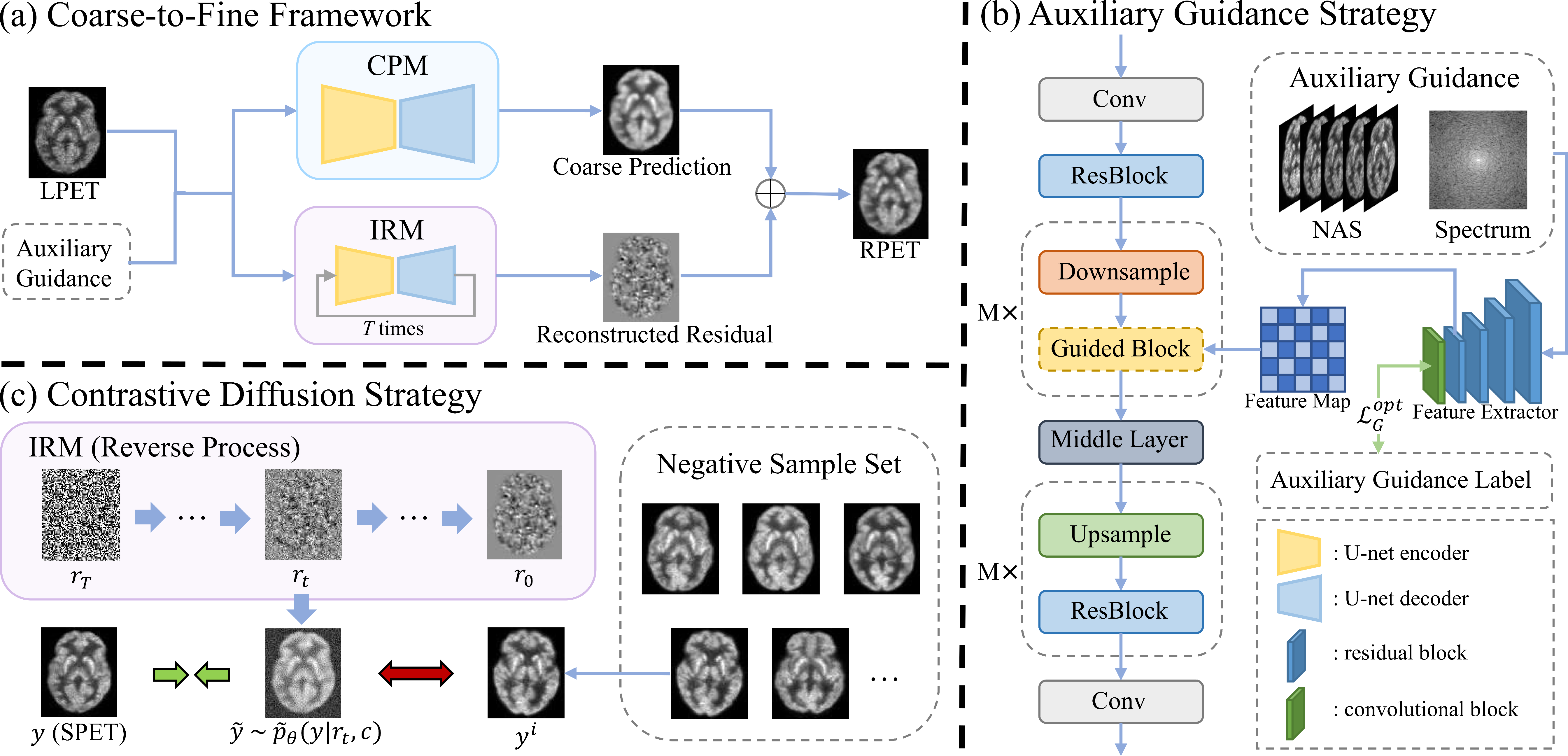}

\caption{Overall architecture of our proposed framework.} \label{fig1}
\end{figure}
\section{Methodology}
Our proposed framework (\figurename~\ref{fig1}(a)) has two modules, i.e., a coarse prediction module (CPM) and an iterative refinement module (IRM). The CPM predicts a coarse-denoised PET image from the LPET image, while the IRM models the residual between the coarse prediction and the SPET image iteratively. By combining the coarse prediction and residual, our framework can effectively generate high-quality RPET images. To improve the correspondence between the LPET image and the RPET image, we adopt an auxiliary guidance strategy (\figurename~\ref{fig1}(b)) at the input level and a contrastive diffusion strategy (\figurename~\ref{fig1}(c)) at the output level. The details of our method are described in the following subsections.

\subsection{Coarse-to-Fine Framework}
To simplify notation, we use a single conditioning variable $c$ to represent the input required by both CPM and IRM, which includes the LPET image $x_{lpet}$ and the auxiliary guidance $x_{aux}$. During inference, CPM first generates a coarse prediction $x_{cp} = {\mathcal{P}}_{\theta}(c)$, where ${\mathcal{P}}_{\theta}$ is the deterministic prediction network in CPM. The IRM, which is the reverse process of DPM, then tries to sample the residual $r_0$ (i.e., $x_0$ in Section 2) between the coarse prediction $x_{cp}$ and the SPET image $y$ via the following iterative process:
\begin{equation}
    r_{t-1}^{\prime} \sim p_{\theta}(r_{t-1}|r_t,c), t = T,T-1, \cdots, 1.
\end{equation}
Herein, the prime symbol above the variable indicates that it is sampled from the reverse process instead of the forward process. When $t=1$, we can obtain the final sampled residual $r_0$, and the RPET image $y^{\prime}$ can be derived by $r_0^{\prime} + x_{cp}$.


In practice, both CPM and IRM use the same network architecture shown in \figurename~\ref{fig1}(c). CPM generates the coarse prediction $x_{cp}$ by using ${\mathcal{P}}_{\theta}$ only once, but the denoising network ${\mathcal{D}}_{\theta}$ in IRM will be invoked multiple times during inference. Therefore, it is rational to delegate more computation overhead to ${\mathcal{P}}_{\theta}$ to obtain better initial results while keeping ${\mathcal{D}}_{\theta}$ small, since the reduction in computation cost in ${\mathcal{D}}_{\theta}$ will be accumulated by multiple times. To this end, we set the channel number in ${\mathcal{P}}_{\theta}$ much larger than that in the denoising network ${\mathcal{D}}_{\theta}$. This leads to a larger network size for ${\mathcal{P}}_{\theta}$ compared to ${\mathcal{D}}_{\theta}$.




\subsection{Auxiliary Guidance Strategy}
In this section, we will describe our auxiliary guidance strategy in depth which is proposed to enhance the reconstruction process at the input level by incorporating two auxiliary guidance, i.e., neighboring axial slices (NAS) and the spectrum. Our findings indicate that incorporating NAS provides insight into the spatial relationship between the current slice and its adjacent slices, while incorporating the spectrum imposes consistency in the frequency domain.

To effectively incorporate these two auxiliary guidances, as illustrated in \figurename~\ref{fig1}(c), we replace the ResBlock in the encoder with a Guided ResBlock as done in \cite{ren2022image}. During inference, the auxiliary guidance $x_{aux}$ is first downsampled by a factor of $2^k$ as $x_{aux}^{k}$, where $k = 1,\cdots,M$, and $M$ is the number of downsampling operations in the U-net encoder. Then $x_{aux}^{k}$ is fed into a feature extractor $\mathcal{F}_{\theta}$ to generate its corresponding feature map $f_{aux}^{k} = \mathcal{F}_{\theta}(x_{aux}^{k})$, which is next injected into the Guided ResBlock matching its resolution through $1 \times 1$ convolution. 

To empower the feature extractor to contain information of its high-quality counterpart $y_{aux}$, we constrain it with $\mathcal{L}_1$ loss through a convolution layer $\mathcal{C}_{\theta}(\cdot)$:
\begin{equation}
    {\mathcal{L}}_{G}^{opt} = \sum_{k=1}^{M} \Vert\mathcal{C}_{\theta}(\mathcal{F}_{\theta}(x_{aux}^{k})) - y_{aux}^{k}\Vert_1,
\end{equation}
where $opt \in $\{{NAS}, {spectrum}\} denotes the kind of auxiliary guidance.
\subsection{Contrastive Diffusion Strategy}
In addition to the auxiliary guidance at the input level, we also develop a contrastive diffusion strategy at the output level to amplify the correspondence between the condition LPET image and the corresponding RPET image. In detail, we introduce a set of negative samples $Neg = \left\{y^1, y^2, ..., y^N\right\}$, which consists of $N$ SPET slices, each from a randomly selected subject that is not in the current batch for training. Then, for the noisy latent residual $r_t$ at time step $t$, we obtain its corresponding intermediate RPET $\widetilde{y}$, and draw it close to the corresponding SPET $y$ while pushing it far from the negative sample $y^i \in Neg$. Before this, we need to estimate the intermediate residual corresponding to $r_t$ firstly, denoted as $\widetilde{r_0}$. According to section 2, the denoising network ${\mathcal{D}}_{\theta}$ manages to predict the Gaussian noise added to $r_0$, enabling us to calculate $\widetilde{r_0}$ directly from $r_t$:
\begin{equation}
    \widetilde{r_0}= \frac{r_t - (\sqrt{1-\gamma_t}){\mathcal{D}}_{\theta}(c, r_t,\gamma_t)}{\sqrt{\gamma_t}}.
\end{equation}

Then $\widetilde{r_0}$ is added to the coarse prediction $x_{cp}$ to obtain the intermediate RPET $\widetilde{y} = x_{cp} + \widetilde{r_0}$. Note that $\widetilde{y}$ is a one-step estimated result rather than the final RPET $y^{\prime}$. Herein, we define a generator $\widetilde{p}_{\theta}(y|r_t,c)$ to represent the above process. Subsequently, the contrastive learning loss ${\mathcal{L}}_{CL}$ is formulated as: 
\begin{equation}
    {\mathcal{L}}_{CL} = \mathbb{E}_{q(y)}[-log \widetilde{p}_{\theta}(y | r_t,c)] - \sum_{y^i \in Neg}\mathbb{E}_{q(y^i)} [-log \widetilde{p}_{ \theta}(y^i | r_t,c)].
\end{equation}




Intuitively, as illustrated in \figurename~\ref{fig1}(b), the ${\mathcal{L}}_{CL}$ aims to minimize the discrepancy between the training label $y$ and the intermediate RPET $\widetilde{y}$ at each time step (first term), while simultaneously ensuring that $\widetilde{y}$ is distinguishable from the negative samples, i.e., the SPET images of other subjects (second term). The contrastive diffusion strategy extends contrastive learning to each time step, which allows LPET images to establish better associations with their corresponding RPET images at different denoising stages, thereby enhancing the mutual information between the LPET and RPET images as done in \cite{zhu2022discrete}.



\subsection{Training Loss}
Following \cite{whang2022deblurring}, we modify the objective ${\mathcal{L}}_{DPM}$ in Eq.\ref{obj}, and train CPM and IRM jointly by minimizing the following loss function:
\begin{equation}
    {\mathcal{L}}_{main} = \mathbb{E}_{(c, y) \sim p_{train}} \mathbb{E}_{\epsilon \sim \mathcal{N} (0, I)} \mathbb{E}_{\gamma \sim p_{\gamma}} \Vert {\mathcal{D}}_{\theta}(c, \sqrt{\gamma} (y-{\mathcal{P}}_{\theta}(c)) + \sqrt{1-\gamma} \epsilon,\gamma) - \epsilon \Vert_1.
\end{equation}

In summary, the final loss function is:
\begin{equation}
    {\mathcal{L}_{total}} = {\mathcal{L}}_{main} + m {\mathcal{L}}_{G}^{NAS} + n {\mathcal{L}}_{G}^{spectrum} + k {\mathcal{L}}_{CL} ,
\end{equation}
where $m$, $n$ and $k$ are the hyper-parameters controlling the weights of each loss.

\subsection{Implementation Details}
The proposed method is implemented by the Pytorch framework using an NVIDIA GeForce RTX 3090 GPU with 24GB memory. The IRM in our framework is built upon the architecture of SR3 \cite{saharia2022image}, a standard conditional DPM. The number of downsampling operations $M$ is 3, and the negative sample set number $N$ is 10. 4 neighboring slices are used as the NAS guidance and the spectrums are obtained through discrete Fourier transform. As for the weights of each loss, we set $m$=$n$=1, and $k$=5$e$-5 following \cite{zhu2022discrete}. We train our model for 500,000 iterations with a batch size of 4, using an Adam optimizer with a learning rate of 1$e$-4. The total diffusion steps $T$ are 2,000 during training and 10 during inference.

\section{Experiments and Results}
\textbf{Datasets and Evaluation: }We conducted most of our low-dose brain PET image reconstruction experiments on a public brain dataset, which is obtained from the Ultra-low Dose PET Imaging Challenge 2022 \cite{Dataset}. Out of the 206 18F-FDG brain PET subjects acquired using a Siemens Biograph Vision Quadra, 170 were utilized for training and 36 for evaluation. Each subject has a resolution of $128\times 128\times 128$, and 2D slices along the z-coordinate were used for training and evaluation. To simulate LPET images, we applied a dose reduction factor of 100 to each SPET image. To quantify the effectiveness of our method, we utilized three common evaluation metrics: the peak signal-to-noise (PSNR), structural similarity index (SSIM), and normalized mean squared error (NMSE). Additionally, we also used an in-house dataset, which was acquired on a Siemens Biograph mMR PET-MR system. This dataset contains PET brain images collected from 16 subjects, where 8 subjects are normal control (NC) and 8 subjects are mild cognitive impairment (MCI). To evaluate the generalizability of our method, all the experiments on this in-house dataset are conducted in a cross-dataset manner, i.e., training exclusively on the public dataset and inferring on the in-house dataset. Furthermore, we perform NC/MCI classification on this dataset as the clinical diagnosis experiment. \textit{\textbf{Please refer to the supplementary materials for the experimental results on the in-house dataset}}.
\begin{table}[]
    \center
    \caption{Quantitative comparison results on the public dataset. *: We implemented this method ourselves as no official implementation was provided.}
    \label{tab:my-table}
    \begin{tabular}{cl|cccc}
    \hline
    \multicolumn{1}{l}{} &  & PSNR↑ & SSIM↑ & NMSE↓ & MParam. \\ \hline
    \multicolumn{1}{c|}{regression-based method} & DeepPET \cite{haggstrom2019deeppet} & 23.078 & 0.937 & 0.087 & \textbf{11.03} \\ \hline
    \multicolumn{1}{c|}{\multirow{4}{*}{GAN-based method}} & Stack-GAN \cite{wang20183d2} & 23.856 & 0.959 & 0.071 & 83.65 \\
    \multicolumn{1}{c|}{} & Ea-GAN \cite{yu2019ea} & 24.096 & 0.962 & 0.064 & 41.83 \\
    \multicolumn{1}{c|}{} & AR-GAN \cite{luo2022adaptive} & 24.313 & 0.961 & 0.055 & 43.27 \\ 
    \multicolumn{1}{c|}{} & 3D CVT-GAN \cite{zeng20223d} & 25.080 & 0.971 & 0.039 & 28.72 \\ \hline
    \multicolumn{1}{c|}{\multirow{3}{*}{likelihood-based method}} & *NVAE \cite{cui2022pet} & 23.629 & 0.956 & 0.064 & 58.24 \\
    \multicolumn{1}{c|}{} & Ours & 25.638 & 0.974 & 0.033 & 34.10 \\
    \multicolumn{1}{c|}{} & Ours-AMS & \textbf{25.876} & \textbf{0.975} & \textbf{0.032} & 34.10 \\ \hline
    \end{tabular}
    \label{tab:comparison}
    \end{table}
\begin{figure}[htb]
    \center
    \centering
    \includegraphics[width=0.9\textwidth]{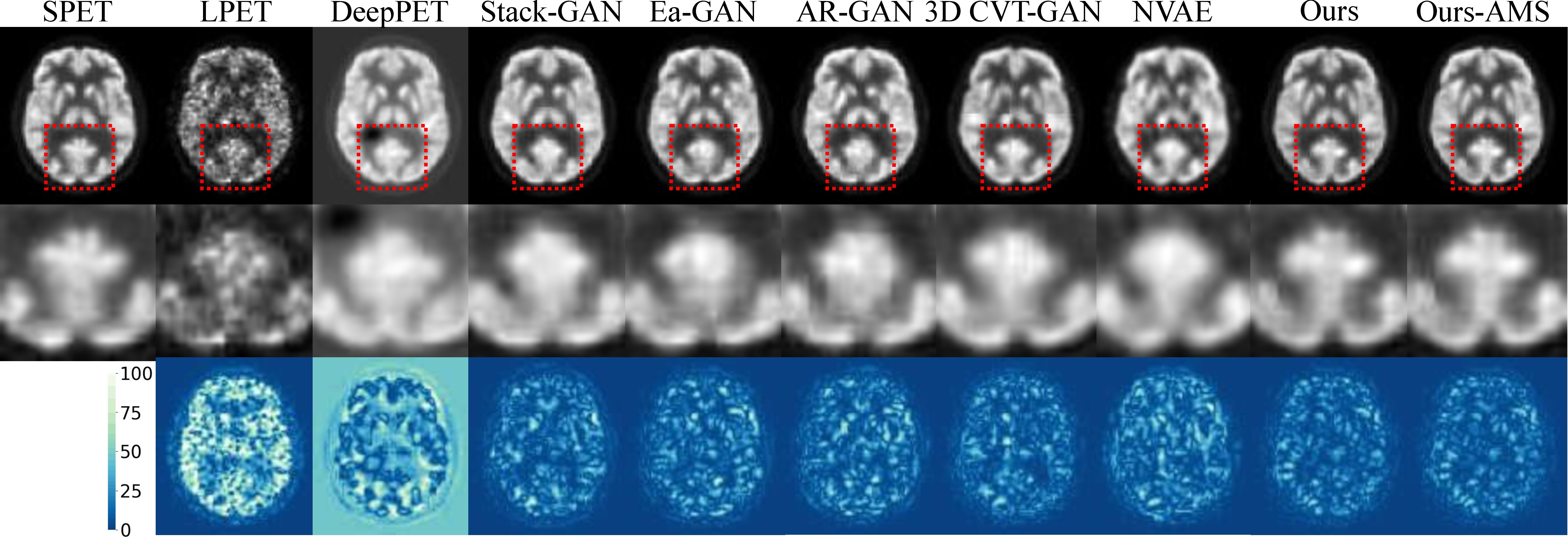}
    \caption{Visual comparison with SOTA methods.}
    \label{fig:comparison}
\end{figure} \\
\textbf{Comparison with SOTA Methods: }We compare the performance of our method with 6 SOTA methods, including DeepPET \cite{haggstrom2019deeppet} (regression-based method), Stack-GAN \cite{wang20183d2}, Ea-GAN \cite{yu2019ea}, AR-GAN \cite{luo2022adaptive}, 3D CVT-GAN \cite{zeng20223d} (GAN-based method) and NVAE \cite{cui2022pet} (likelihood-based method) on the public dataset. Since the IRM contains a stochastic process, we can also average multiple sampled (AMS) results to obtain a more stable reconstruction, which is denoted as Ours-AMS. Results are provided in \tablename~\ref{tab:comparison}. As can be seen, our method significantly outperforms all other methods in terms of PSNR, SSIM, and NMSE, and the performance can be further amplified by averaging multiple samples. Specifically, compared with the current SOTA method 3D CVT-GAN, our method (or ours-AMS) significantly boosts the performance by 0.558dB (or 0.796dB) in terms PSNR, 0.003 (or 0.004) in terms of SSIM, and 0.006 (or 0.007) in terms of NMSE. Moreover, 3D CVT-GAN uses 3D PET images as input. Since 3D PET images contain much more information than 2D PET images, our method has greater potential for improvement when using 3D PET images as input. Visualization results are illustrated in \figurename~\ref{fig:comparison}. Columns from left to right show the SPET, LPET, and RPET results output by different methods. Rows from top to bottom display the reconstructed results, zoom-in details, and error maps. As can be seen, our method generates the lowest error map while the details are well-preserved, consistent with the quantitative results. 
\begin{table}[]
    \caption{Quantitative results of the ablation study on the public dataset.}
    \label{tab:my-table}
    \center
    \begin{tabular}{l|ccccc|cccc}
    \hline
     & \multicolumn{5}{c|}{Single Sampling} & \multicolumn{4}{c}{Averaged Multiple Sampling} \\ \cline{2-10} 
     & PSNR↑ & SSIM↑ & NMSE↓ & MParam. & BFLOPs & PSNR↑ & SSIM↑ & NMSE↓ & SD \\ \hline
    (a)baseline & 23.302 & 0.962 & 0.058 & 128.740 & 5973 & 23.850 & 0.968 & 0.052 & 6.16e-3 \\
    (b)CPM & 24.354 & 0.963 & 0.049 & \textbf{24.740} & \textbf{38} & - & - & - & - \\
    (c)+IRM & 24.015 & 0.966 & 0.044 & 31.020 & 132 & 24.339 & 0.967 & 0.041 & 3.78e-3 \\
    (d)+NAS & 24.668 & 0.969 & 0.046 & 33.040 & 140 & 24.752 & 0.970 & 0.044 & 3.41e-3 \\
    (e)+spec. & 25.208 & 0.972 & 0.044 & 34.100 & 145 & 25.376 & 0.973 & 0.043 & 3.30e-3 \\
    (f)+$\mathcal{L}_{CL}$ & \textbf{25.638} & \textbf{0.974} & \textbf{0.033} & 34.100 & 145 & \textbf{25.876} & \textbf{0.975} & \textbf{0.032} & \textbf{2.49e-3} \\ \hline
    \end{tabular}
    \label{tab:ablation}
\end{table}\\
\textbf{Ablation Study: }To thoroughly evaluate the impact of each component in our method, we perform an ablation study on the public dataset by breaking down our model into several submodels. We begin by training the SR3 model as our baseline (a). Then, we train a single CPM with an L2 loss (b), followed by the incorporation of the IRM to calculate the residual (c), and the addition of the auxiliary NAS guidance (d), the spectrum guidance (e), and the ${\mathcal{L}}_{CL}$ loss term (f). Quantitative results are presented in \tablename~\ref{tab:ablation}. By comparing the results of (a) and (c), we observe that our coarse-to-fine design can significantly reduce the computational overhead of DPMs by decreasing MParam from 128.740 to 31.020 and BFLOPs from 5973 to 132, while achieving better results. The residual generated in (c) also helps to improve the result of the CPM in (b), leading to more accurate PET images. Moreover, our proposed auxiliary guidance strategy and contrastive learning strategy further improve the reconstruction quality, as seen by the increase in PSNR, SSIM, and NMSE scores from (d) to (f). Additionally, we calculate the standard deviation (SD) of the averaged multiple sampling results to measure the input-output correspondence. The standard deviation (SD) of (c) (6.16e-03) is smaller compared to (a) (3.78e-03). This is because a coarse RPET has been generated by the deterministic process. As such, the stochastic process IRM only needs to generate the residual, resulting in less output variability. Then, the SD continues to decrease (3.78e-03 to 2.49e-03) as we incorporate more components into the model, demonstrating the improved input-output correspondence. 

\section{Conclusion}
In this paper, we propose a DPM-based PET reconstruction framework to reconstruct high-quality SPET images from LPET images. The coarse-to-fine design of our framework can significantly reduce the computational overhead of DPMs while achieving improved reconstruction results. Additionally, two strategies, i.e., the auxiliary guidance strategy and the contrastive diffusion strategy, are proposed to enhance the correspondence between the input and output, further improving clinical reliability. Extensive experiments on both public and private datasets demonstrate the effectiveness of our method.

\subsubsection{Acknowledgements.} This work is supported by National Natural Science Foundation of China (NSFC 62071314), Sichuan Science and Technology Program 2023YFG0263, 2023YFG0025, 2023NSFSC0497.

\bibliographystyle{splncs04}
\bibliography{refs}

%





%
\title{Supplementary Material}
%
%
\author{}
%
%
\institute{}
\maketitle              
%

%
%
%
\section{Experimental Results on the In-house Dataset Using Cross-dataset Setup}


%
\begin{table}[H]
    \centering
    \caption{Quantitative results compared with SOTA methods on the in-house dataset using a cross-dataset experimental setup. All models in comparison are trained using the public dataset mentioned in the full paper and directly applied on the in-house dataset without finetuning. Our approach achieves the best result in terms of PSNR, and the second-best result in terms of SSIM and NMSE with negligible lags. * indicates our own implementation of this method as no official code was provided.}
    \begin{tabular}{c|l|ccc}
    \hline
    \multicolumn{1}{l|}{} &  & PSNR↑ & SSIM↑ & NMSE↓ \\ \hline
    regression-based method & DeepPET & 7.452 & 0.904 & 0.676 \\ \hline
    \multirow{4}{*}{GAN-based method} & Stack-GAN & 11.274 & 0.945 & 0.292 \\
     & Ea-GAN & 17.079 & 0.971 & 0.075 \\
     & AR-GAN & 16.470 & 0.968 & 0.087 \\
     & 3D CVT-GAN & 20.241 & \textbf{0.985} & \textbf{0.032} \\ \hline
    \multirow{3}{*}{likelihood-based method} & *NVAE & 19.126 & 0.979 & 0.047 \\
     & Ours & 20.409 & 0.983 & 0.035 \\
     & Ours-AMS & \textbf{20.536} & 0.983 & 0.034 \\ \hline
    \end{tabular}
    \end{table}

\begin{figure}[H]
    \centering
\includegraphics[width=\textwidth]{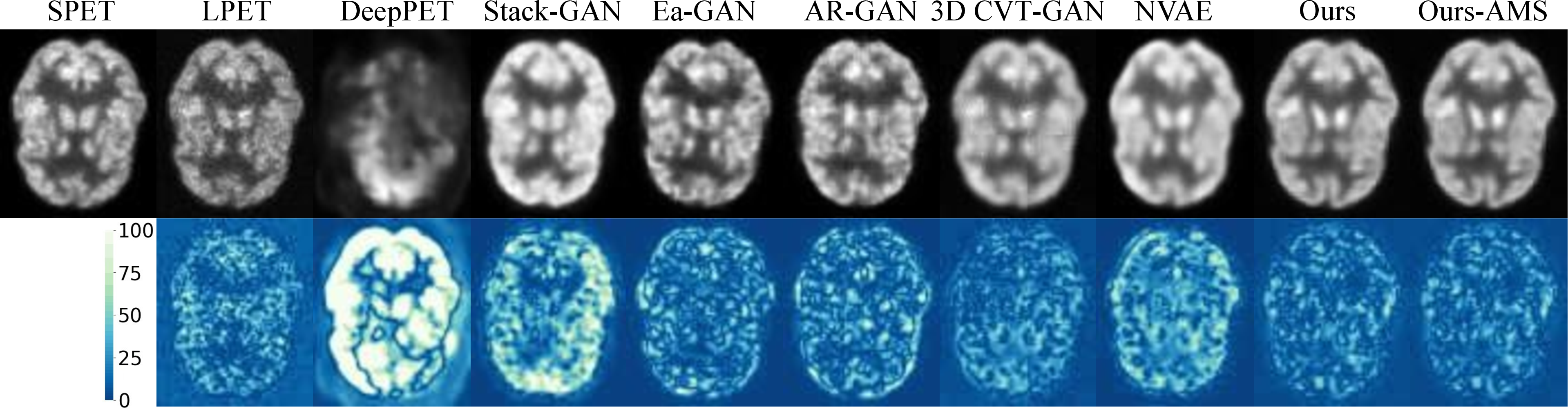}
\caption{Qualitative comparison on the in-house dataset using a cross-dataset experimental setup. The visual results demonstrate the high generalizability of our approach, which generates high-quality RPET images with the lowest residual error on an unseen dataset. Conversely, DeepPET method tends to overfit to the training dataset, leading to corrupted images on the in-house dataset. These results confirm the superior performance of our method in the challenging task of cross-dataset PET reconstruction.}
\end{figure}

\begin{figure}[H]
    \centering
\includegraphics[width=0.6\textwidth]{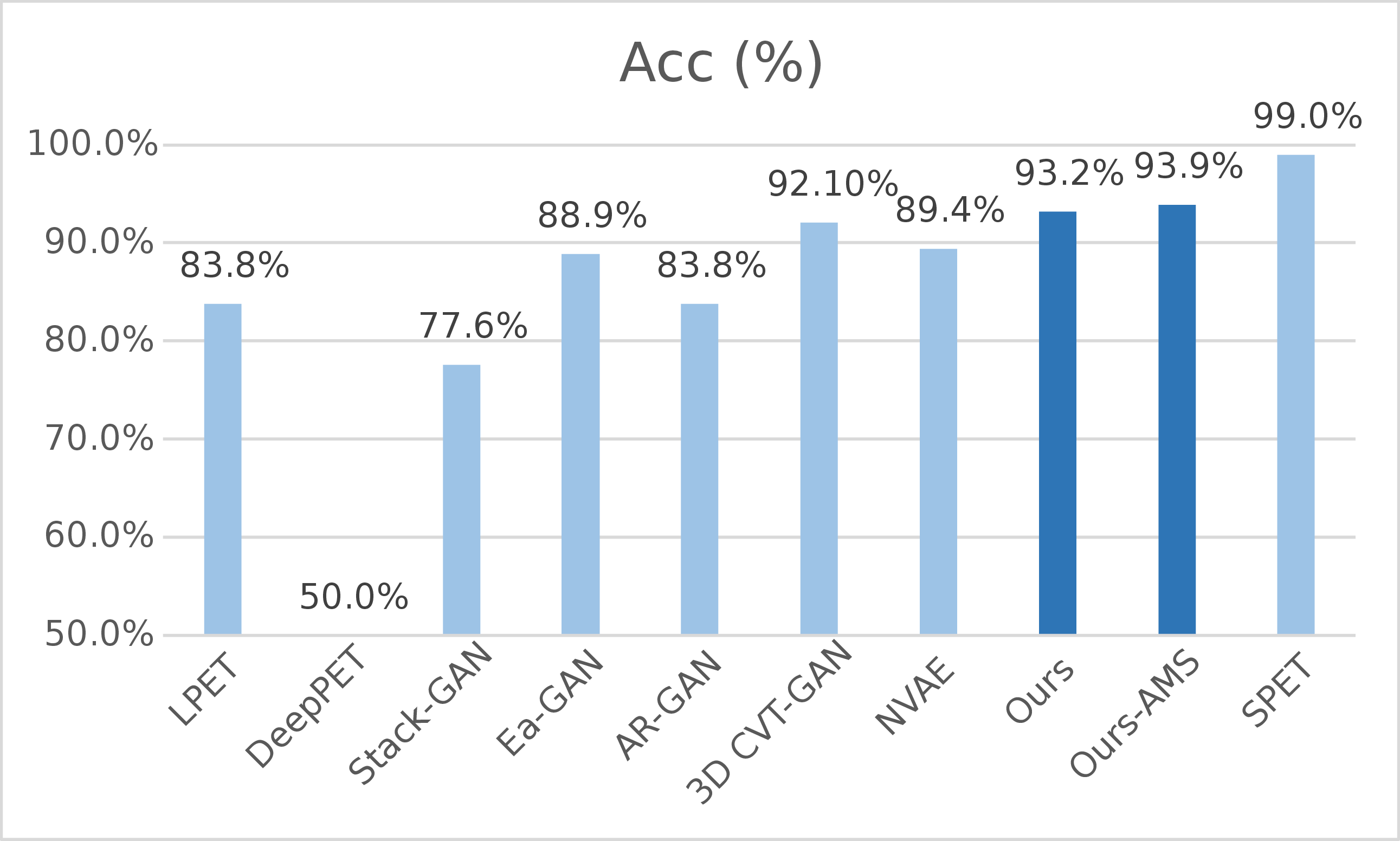}
\caption{Results of clinical NC/MCI diagnosis using a cross-dataset experimental setup. In particular, we trained a multi-layer CNN classification network on SPET images to differentiate NC and MCI subjects. We then employed PET images reconstructed by different methods for testing. Our evaluation demonstrates that our method achieves the highest clinical reliability, as it generates RPET images better maintaining clinic relevant information.}
\end{figure}

\end{document}